# PROTOSTELLAR JETS: THE BEST LABORATORIES TO INVESTIGATE ASTROPHYSICAL JETS

**Elisabete M. de Gouveia Dal Pino**

University of São Paulo, Instituto Astronômico e Geofísico, Av. Miguel Stéfano 4200, São Paulo, SP 04301-904, Brazil

## ABSTRACT

Highly collimated supersonic jets are observed to emerge from a wide variety of astrophysical objects, ranging from Active Nuclei of Galaxies (AGNs) to Young Stellar Objects (YSOs) within our own Galaxy. Despite their different physical scales (in size, velocity, and amount of energy transported), they have strong morphological similarities. Thanks to the proximity and relatively small timescales, which permit direct observations of evolutionary changes, YSO jets are, perhaps, the best laboratories for cosmic jet investigation. In this lecture, the formation, structure, and evolution of the YSO jets are reviewed with the help of observational information, MHD and purely hydrodynamical modeling, and numerical simulations. Possible applications of the models to AGN jets are also addressed.

## 1. INTRODUCTION

Supersonic collimated outflows are an ubiquitious phenomenun in the universe. They span a large range of luminosity and degree of collimation, from the most powerful examples observed to emerge from the nuclei of active galaxies (or AGNs) to the jets associated to low-mass Young Stellar Objects (YSOs) within our own Galaxy. In the intermediate scales between these two extremes one finds evidences of outflows associated to neutron stars, binary systems (with SS433 being the best example of this class), and symbiotic novae. Also, there is some evidence of a poorly collimated jet associated to our Galactic Centre (e.g., [1]).

As stressed by Königl [2], both the AGN and the YSO jets, despite their different physical scales (AGN jets have typical sizes $\geq 10^6$ pc (1 pc = 3.086 $10^{18}$ cm), nuclear velocities ~ c (where c is the light speed), and parent source masses $10^8$ $M_\odot$ ($M_\odot$ = one solar mass = 1.99 $10^{33}$ g), while YSO jets have typical sizes < 1 pc, nuclear velocities $\leq 10^{-3}$c, and central masses ~ 1 $M_\odot$), are morphologically very similar. In both classes, the jets are: (i) highly collimated and in most cases two-sided; (ii) originate in compact objects; (iii) show a chain of more or less regularly spaced emission knots which in some cases move at high speeds away from the central source; (iv) terminate in emission lobes (with line emission in the case of the YSOs and synchrotron continuum emission in the case of the AGN jets), which are believed to be the "working surfaces" where the jets shock against the ambient medium (see below); (v) are associated with magnetic fields whose projected directions are inferred from polarization measurements (in YSOs, these vectors are generally aligned with the jet axis); and (vi) are possibly associated to accretion disks. For comparison, Figs. 1 and 2 show some of the finest examples of YSO (Fig. 1) and AGN (Fig. 2) jets.

Extensive reviews of the observational properties of the YSO jets can be found in [1, 3, 4, 5, 6, 7, 8, 9]. Reviews of the properties of the AGN jets can be found in [e.g., 10, 11, 12]. Jets from young stars offer to investigators a number of advantages relative to their extragalactic counterparts. They are nearby and generally bright, and evolve on relatively short timescales, permitting direct observations of evolutionary changes. Furthermore, their radiation which is mainly line-emission rather than continuum, opens a vast field of observational techniques that can probe their physical conditions [e.g., 5]. Therefore, from the YSO jets we can potentially learn a great deal about the structure, evolution and driving mechanisms of the astrophysical jets as a hole.

In this work, instead of discussing all the current knowledge about the various classes of astrophysical jets (which would be impossible to cover just in few pages), I focuse on the YSO jets which are the closest and therefore, the best laboratories for cosmic jet investigation. I do not address here specific issues in the theory of AGN jets, which have been the subject of several extensive reviews [e.g., 10, 11, 12, 13, 14, 15].

The review begins with a brief description of the observed characteristics of the YSO jets (Section 2). A discussion of the basic structural components of the jet and the analytical and numerical models which explain these features is presented in Sections 3 and 4 giving particular emphasis to the study of the jet head structure (Section 3) and the possible mechanisms for the formation of the internal knots along the jets (Section 4). The momentum transfer processes between jet and environment and the formation of molecular outflows driven by the YSO jets are addressed in Section 5. In Section 6, the effects of the propagation of the jets in non-

**Fig. 1.** Contour plot of a [SII] image of the HH 34 system. The inset in the lower pannel is a 4 × amplification of the jet with its chain of aligned knots. The arrows give the motions of the individual knots. The bow shaped condensations at the edges are also visible (from S. Heathcote and B. Reipurth 1992, Astron. J., 104, 2193, see also [5]).

**Fig. 2.** Radio map of the extragalactic jet at the frequency MHz. As in Fig. 1, the knots and the lobes at the edge of the system are clearly visible (from homogeneous stratified environments are examined and in Section 7 the existing mechansims for the origin of the jets from the central YSO object and associated accretion disk is discussed. A summary is presented in Section 8.

## 2. OBSERVATIONAL PROPERTIES OF YSO JETS

The highly collimated supersonic YSO jets have typical projected lengths between ~0.01-1 pc. Most of them show a linear chain of bright, traveling knots, frequently identified as Herbig-Haro (HH) objects (e.g., HH34, and HH111 systems), followed by an intermediate section where the emission disappears, and terminating in a bow shock-like structure identified as the working surface where the jet impacts with a slower ambient gas (Fig. 1) [e.g., 5]. In some cases (e.g., HH46/47 and HH 111 jets), there is clear evidence of two or more such bow shock structures separated by a trail of diffuse gas for many jet radii. In other cases (e.g., HH30 jet), there is no detected bow-shock like feature [8]. Some jets, instead of chains of aligned knots, show larger amplitude side-to-side "wiggles" (e.g., HH83 and HH110 jets; [16]).

All the luminous structures produce emission line spectra. Prominent features include the hydrogen Balmer lines and transitions of neutral atoms ([OI], [CI], [NI]) and ions (CaII, [CaII], [FeII], [SII], [OIII]). While the terminal feature may be rich in high ionization and excitation lines, the inner jet generally shows a very low excitation spectrum with [SII] and [OII] to H$\alpha$ ratios substantially greater than unity [5, 16]. This implies that the temperature of the YSO jets are not much larger than $\approx 1\text{-}2 \times 10^4$ K and the corresponding sound speeds are ~10 km/s. This yields typical Mach numbers for the emitting regions $M_j$=10-30 [8].

The knots move away from the sources at high speeds ~100-500 km/s, have radii $R_j \approx 3 \times 10^{15}$ cm, and inter-knot separation $\Delta x \approx 10^{16}$ cm $\approx 3.3\ R_j$. They often become fainter and disappear at larger distances from the source (e.g., HH34 and HH 111 jets; [5, 8]).

The densities of the YSO jets are poorly determined. Assuming that the line emission is produced by shock-heated gas, early estimates indicated jet number densities $n_j \approx 20\text{-}100$ cm$^{-3}$ [3], but recent estimates seem to indicate values larger than $10^3$ cm$^{-3}$ [e.g., 18, 19, 20, 21].

The high proper motion of the heads of the jets indicate that their density is considerably higher than the density of the surrounding medium. Although uncertain, observations suggest a jet-to-environment density ratio $\eta = n_j/n_a \approx 1\text{-}20$ [e.g., 3, 18, 20].

The typical jet velocities and total lengths imply dynamical timescales $L_j/v_j$ ~1000 yrs or larger for the jets.

The determination of the intensity of the magnetic field in stellar jets from line intensity ratios is very imprecise because it is done under the assumption that the lines are produced behind radiative shocks and the emisson line ratios from a radiative shock with a magnetic field resemble those of a lower velocity shock without a magnetic field [21]. However, estimates for the HH34 jet [18] indicate a magnetic field = 8 $\mu$G in front of HH34 head.

## 3. THE JET STRUCTURE

The basic theoretical properties of adiabatic and cooling jets have been summarized by Norman [14] and Blodin et al. [22]. A supersonic jet propagating into a stationary ambient gas will develop a shock pattern at its head also denominated working surface. While the impacted ambient material is accelerated by a forward bow shock, the beam outflow is decelerated in a jet shock or Mach disk. The velocity of advance of the bow shock can be estimated by balancing the momentum flux of the jet material at the jet head with the momentum flux generated by the ambient medium at the bow shock. For highly supersonic flows, the thermal pressure of the jet and ambient medium can be neglected and one obtains:

$$v_{bs} \approx v_j\ [1 + (\eta\alpha)^{-1/2}]^{-1} \qquad (1)$$

where $v_j$ is the jet velocity, $v_{bs}$ is the velocity of advance of the bow shock, $\eta = n_j/n_a$ is the ratio of the jet number density to the ambient number density, and $\alpha = (R_j/R_h)^2$, where $R_j$ is the radius of the jet beam and $R_h$ is the radius at the jet head. From eq. (1) we see that for a low-density jet ($\eta \ll 1$) which is generally believed to be the case for AGN jets, $v_{bs} \ll v_j$ and the jet material is constantly decelerated at the end of the jet (at the jet shock). The high-pressure, shocked gas drives a flow back around the jet forming a cocoon of shock-heated waste material. A dense jet, like typical YSO jets, on the other hand, for which $v_{bs} \approx v_j$ ($\eta \gg 1$), will simply plow through the ambient medium at close to the jet velocity without accumulating much waste of gas in the cocoon. The ambient material that traverses the bow shock will form a shroud of ambient shock-heated gas surrounding the beam/cocoon structure. Fig. 3 illustrates the basic components of a supersonic jet.

**Fig. 3.** The basic structural features of a jet: the beam, the working surface (with the bow shock and the jet shock), the cocoon, and the surrounding shroud (from Blodin et al. [22]).

The jets of AGNs were the focus of much theoretical work during the last decade. Extensive numerical simulations were performed under the *adiabatic* approach (e.g., [14] and references therein) in order to study their structure, stability properties, and propagation through the ambient medium. Considering, however, the typical number densities and shock velocities of the YSO jets one can evaluate the radiative cooling time of the shock-heated material:

$$t_{cool} = (n_e + n_H) k T_s / [(\gamma-1) n_e n_H \Lambda(T_s)] \quad (2)$$

where $n_e n_H \Lambda(T_s)$ is the cooling rate in units of energy per unit volume and time, $n_e$ and $n_H$ are the postshock electronic and hydrogen number densities, respectively, $\gamma$ is the ratio of specific heats of the gas, k is the Boltzmann constant, and $T_s$ is the immediate postshock temperature $T_s = 2(\gamma-1) \mu v_s^2 / [(\gamma+1)^2 k]$, where $\mu$ is the average mass per particle. Taking shock velocities $v_s \sim 40\text{-}100$ km/s, inferred from the observed line-ratios of the emitting matter in the knots, and number densities $\approx 100$ cm-3 one finds tcool $\approx 100$ yrs, which is, in general, smaller than the evolutionary time of these objects. The assumption of an adiabatic gas is therefore inappropriate for YSO jets.

The importance of the cooling can be also quantified in terms of the cooling length, $d_{cool}$, for the gas to cool to some low value ($\approx 10^4$ K) behind one shock, $d_{cool} \approx v_s t_{cool}/4$ (where vs/4 is the postshock speed for a strong shock). Thus, defining the cooling parameter $q_s = d_{cool}/R_j$, one expects that if $q_s < 1$, the shock is fully radiative and the postshock gas loses the bulk of its internal energy in a relatively short distance. In this case, the jet develops a shell of dense gas formed by the cooling of the shock-heated gas in the working surface. Considering the radiative cooling function $\Lambda(T)$ (due to collisional excitation and recombination) for a cosmic abundant gas, Blodin et al. [22] evaluated $q_s$ for a gas cooling from T $\approx 10^6$ to 8000 K and found that for $v_{s,7} \geq 0.9$ (where $v_{s,7}$ is the shock velocity in units of $10^7$ cm/s) the cooling parameter for the bow shock is

$$q_{bs} = d_{cool}/R_j \approx 4 \times 10^{16} n_o^{-1} R_j^{-1} v_{j,7}^4 [1 + (\eta\alpha)^{-1/2}]^{-4} \quad (3)$$

where $n_o$ is the ambient number density of nuclei, and noting that the jet shock velocity is given by $v_{js} = v_j - v_{bs}$, the cooling parameter for the jet shock is

$$q_{js} \approx q_{bs} \eta^{-3} \alpha^{-2} \quad (4)$$

Assuming $\alpha \approx 1$, these relations imply that a heavy jet ($\eta \gg 1$) will exhibit much stronger cooling behind the jet shock and most of the gas that accumulates in the shell in this case consists of shocked jet material. The reverse situation holds for light ($\eta \ll 1$) jets.

The analytic predictions of the equations above are confirmed by multidimensional simulations [22, 24, 25]. Hydrodynamical models of cooling jets are characterized by the dimensionless parameters: $\eta$ (eq. 1), $M_a = v_j/c_a$ the ambient Mach number, where $c_a$ is the ambient sound speed; and $q_{bs}$ (eq. 3). As an example, Fig. 4 shows the results of a fully three-dimensional hydrodynamical simulation of a jet (ejected into the ambient medium with continuous velocity) from Gouveia Dal Pino and Benz [24] which was performed with the employment of the smoothed particle hydrodynamics technique (SPH).

In Fig. 4, a dense shell has developed in the head of the jet formed by the cooling of the shock-heated gas in the working surface. In this case, $q_{js} \approx 0.39$ and $q_{bs} \approx 10.5$, implying that a radiative jet ($q_{js} < 1$) is propagating into an adiabatic ambient medium ($q_{bs} \gg 1$) and the gas that accumulates in the dense shell consists essentially of shocked jet material. The little cooling of the ambient gas, results a high pressure shroud which helps to confine the beam and the cocoon. The shell is responsible for most of the emission of the jet. In the figure it has become dynamically (Rayleigh-Taylor) unstable and fragmented into pieces that have spilled out to the cocoon forming, together with the shell, an elongated plug of cold gas in the head. The fragmented shell resembles the clumpy structure observed in many HH objects at the bow shock of YSO jets (e.g., HH1, HH2, HH19, HH12). The fragmentation is accompanied by density variations with time. These variations are associated to global thermal instabilities which cause the radiative shock front to move back and forth with respect to the shell and oscillate between a quasi-adiabatic and a radiative shock regime. These oscillations may explain the brightness variability observed in objects like HH1 and HH2.

**Fig. 4.** The central density contours of a cooling jet with $\eta=3$, $n_j=60$ cm$^{-3}$, $R_j=2 \times 10^{16}$ cm, $v_j=400$ km/s, initial $v_{bs} \approx 254$ km/s and $v_{js} \approx 146$ km/s, $M_a=11.55$, $M_j=v_j/c_j=20$. The contour lines of the gas density are separated by a factor of 1.3 and the density scale covers the range from $\approx 0.05$ up to $\approx 1400/n_a$. The entire evolution corresponds to an age $\approx 782$ yrs (from Gouveia Dal Pino and Benz, 1993 [24]).

Although of fundamental importance for the production and initial collimation of the jet (see Section 7), the magnetic field effects have been neglected in the calculations above and, in fact, most of the modeling of the *structure and evolution* of YSO jets has been made without taking into account the magnetic field (see also Sections 4, 5, and 6). The determination of its intensity from observations is very imprecise but rough estimates for HH34 region [18], for example, indicate a magnetic field Å 8 µG in front of HH34 (see Section 2) with an energy density in the preshock gas $<10^{-3}$ of the ram pressure ($\rho_j v_j^2$), implying that the magnetic field is not relevant in the dynamics of the HH34, at least in the undisturbed medium. However, the magnetic field may become important once it has been amplified by compression behind a radiative shock. Its presence can, for example, inhibit the maximum density amplification of the cold shell in the cooling jet. A microgauss field, compressed a hundredfold in the shell is sufficient to begin to stabilize the shell against the Rayleigh-Taylor instability [22] and a field strength, transverse to the flow, of 9 mG is sufficient to suppress the global oscillatory thermal instability in a shock with a velocity of 175 km/s, traveling into a medium with a density of 1 $cm^{-3}$. In smaller velocity shocks, the oscillations are damped with even smaller field strengths [26]. In the future, it will be fruitful to incorporate magnetic fields in the numerical simulations of radiative cooling jets in order to examine their effects on the structure of the flow, particularly at the shocked gas. A first attempt in this way has been recently done by for bow shock models [27].

## 4. THE ORIGIN OF THE KNOTS

The precise nature of the mechanism that produces the internal knots is still controversial. Most of the available models rely on the belief that the knots represent radiative shocks excited within a jet flow.

Among the proposed models analytic and numerical calculations show that a chain of stationary oblique crossing shocks, such as those observed in laboratory jets, is produced when an initially freely expanding jet attempts to reach pressure equilibrium with its surroundings [e.g., 28, 29]. When a nearly isotropic wind blown by the star perforates the circumstellar flattened envelope in the polar directions a "De Laval nozzle" is produced which focuses the flow. Passing through the throat of the nozzle the flow develops a chain of stationary oblique crossing shocks, similar to those observed in laboratory jets, that have the function of adjusting the flow to the ambient pressure. However, the fact that the knots are observed to move away from the source at high speeds seems to exclude this mechanism since it necessarily produces stationary shocks.

Kelvin-Helmholtz (K-H) shear instabilities at the boundary between the jet and the surrounding medium may also excite a similar pattern of oblique internal shocks (e.g., [30-33]). This mechanism produces shocks which travel at a velocity $v_p$ that can be substantially smaller than the fluid velocity $v_j$. Previous 2-D simulations of steady-state, adiabatic jets by Norman et al. [31] show that $v_p/v_j$ increases with both increasing Mach number and increasing jet-to-environment density ratio η, approaching the flow velocity at the limit of a highly supersonic heavy jet, as is the case of the protostellar jets. However, recent two-dimensional [22] and three-dimensional [24] hydrodynamic simulations of radiative cooling jets have shown that, under the presence of radiative cooling, shocks driven by K-H instabilities are fainter and less numerous than in adiabatic jets. The radiative cooling reduces the thermal pressure which is deposited in the cocoon that surrounds the beam. As a result, the cocoon has less pressure to collimate, drive K-H instabilities and thus, reflect internal shocks in the beam. In Gouveia Dal Pino & Benz [24] simulations, only the very heavy cooling jet (η =10) shows the formation of internal shocks due to K-H instability, but these turn out to be stronger, and consequently the Hα emission higher, near the jet head than closer to the star, contrary to what is usually observed. Although this mechanism remains viable, it probably plays a secondary role in the production of most of the internal shocks, at least in the case of the YSO jets for which the radiative cooling is important (see below, however, the formation of knots by K-H instabilities in jets propagating in stratified ambients of increasing density; Section 6). In the case of the extragalactic jets, for which the adiabatic approach is good, the simulation depicted in Fig. 5 shows that the development of the pinch and helical modes of the K-H instability can be relevant in the production of internal knots.

Time variations in the ejection mechanism that produces the jet outflow may also create traveling internal shock patterns similar to the *working surfaces* at the edges of the jets [e.g., 34-42] (For a review of the properties of models of jet flows from sources with variability in the ejection velocity and/or direction see also [8].) Strong support for this mechanism comes from recent high resolution observations of the HH34 (Fig. 1) and HH111 jets [18] which show evidence of a velocity-variable outflow with multiple ejections. In the case of HH46/47 jet, observed abrupt variations of the radial velocity (that is, the velocity along the line of sight) are also consistent with a variable jet ejection velocity mechanism [5]. The time variability of the outflow is probably associated with irruptive events in the accretion process around the protostar (see Section 7). Some energy sources of YSO jets and HH objects (e.g., HH57 and HH28/29), have been seen to erupt into FU Orionis outbursts [16]. These episodes could produce massive ejections along the stellar jets every 100-1000 yrs [37], suggesting that the driving sources can change on timescales shorter than the dynamical timescale of the HH objects.

Numerical simulations of jets from time-dependent sources clearly show the formation of internal shocks [37-41]. Fully 3-D numerical SPH simulations of cooling jets with time variable ejection velocity performed by Gouveia Dal Pino and Benz [39] are shown in Fig. 6. To produce the velocity variability, they assumed that the jet is periodically turned on with a full supersonic velocity and periodically turned off to a small velocity regime. Considering velocity variations with a period comparable to the transverse dynamical time scale of the jet ($\tau_{dy} \equiv R_j/v_j$, where $R_j$ is the jet radius and $v_j$ is the mean jet velocity), they found the chain of regularly spaced emitting features depicted in Fig. 6. Each parcel in Fig. 6 develops a pair of shocks: a forward (downstream) shock which sweeps up the low velocity material ahead of it and propagates downstream in the jet with a velocity $v_{is}$ (eq. 5, see below); and a reverse (upstream) shock which decelerates the high velocity material behind it and propagates with a velocity $v_{rs} \approx v_j - v_{is}$. One-dimensional analysis gives an estimate of the velocity of propagation of an internal working surface. For highly supersonic flows it is given by

$$v_{is} \approx (\beta v_u + v_d)/(1+\beta) \qquad (5)$$

where $\beta = (n_u/n_d)^{1/2}$, $n_u$ and $n_d$ are the number densities up and downstream of the working surface, $v_u$ is the velocity upstream and $v_d$ the velocity downstream. We see that for $\beta \gg 1$

$$v_{is} \approx (1 - \beta^{-1}) v_u \qquad (6)$$

**Fig. 5** Density contours and velocity distribution of an adiabatic jet propagating in an initially homogeneous ambient medium with density ratio $\eta=1$ and a Mach number $M_a=v_j/c_a=20$. The evolution corresponds to a jet age $\approx 1115$ yrs. Thanks to the high pressure cocoon of the adiabatic jet, the beam becomes K-H unstable with the appearance of reflecting pinch and helical modes which collimate the beam and drive internal shocks that propagate downstream with velocities of the order of the propagation speed of the jet head (from Gouveia Dal Pino and Benz 1993 [24]).

The simulations show that the features (parcels) move downstream with nearly the jet velocity in the supersonic phase $v_{is} \approx v_j = 150$ km/s. This result is in agreement with the observed knots of HH34 jet which seem to move at a velocity very close to the flow speed. This result is also consistent with eqs. (5) and (6) for a density ratio of the fast material (upstream of the supersonic discontinuity) to the slow material (downstream of the discontinuity) $\beta^2 \gg 1$. As a consequence, the propagation velocity of the reverse shock $v_{rs} \approx 0$ and the forward shock (of each parcel) is much stronger than the reverse shock and line emission between these shocks is essentially single peaked. Also consistent with the observations is the knot separation of the order of few jet radii ($\Delta x \approx 5 R_j$). As the peaks travel downstream, they widen and fade and eventually disappear close to the leading working surface, an effect also detected in HH 34 jet. This effect is due to the increase in the pressure of the postshock gas which causes both the separation between each pair of shocks and the expulsion of jet material laterally to the cocoon and to the rarefied portions of the beam itself. This fading explains the most frequent occurence of knots closer to the driving source. The maximum density behind the internal shocks is smaller than the maximum density at the head of the jet. This implies that the emission from internal shocks is of smaller intensity and excitation than the emission from the head, in agreement with the observations. These results are also consistent with previous one and two-dimensional˝numerical calculations [37, 38, 40-42].

**Fig. 6** Intermittent jet periodically turned on with a supersonic velocity $v_j$=150 km/s and periodically turned off with a subsonic velocity 15 km/s. The *turning on* and quiescent periods are both given by $\tau_{on} = \tau_{off} = (R_j/c_a) \approx$ 127 yrs = 3 $\tau_{dy}$. The initial input parameters are $\eta = 10$, $n_a$=1000 cm$^{-3}$, $R_j = 2 \times 10^{16}$ cm, and $M_a = 3$. The entire evolution corresponds to t $\approx$ 11 $R_j/c_a$ = 1387 yr (from Gouveia Dal Pino and Benz 1994 [39]).

**Fig. 7** Central density contours and the corresponding velocity distribution of an $\eta = 3$ cooling jet with a long-period variability. The turned on period is $\tau_{on} = 22.5\ \tau_{dy} = 143$ yr with a Mach number $M_a = 17.3$ and the turned off period is $\tau_{off} = 17.3\ \tau_{dy}$ with a weakly supersonic velocity. The calculation ends up at $t'' \approx 587\ (R_j/v_j) \approx 374$ yr ($z_{max} \approx 45\ R_j$). A pair of bow shaped working surfaces separated by a long trail of very diffuse gas have been produced. They resemble the pair of bow-shock like features observed in systems like the HH111 jet (from Gouveia Dal Pino and Benz 1994 [39]).

The fact that the time dependent jet model successfully explains a number of properties of the knots seems to favor it over the K-H instability model. However, a way to distinguish between both could come from detailed kinematic studies. In an attempt to do that, Eislofel & Mundt [43], estimating an angle $\theta = 23^o$ between the HH34 jet and the plane of the sky, tried to evaluate the ratio $v_p/v_j = v_t/v_r\ \tan\theta$ directly from the observed tangential (vt) and radial velocities (vr) of the knots. They found that $v_p/v_j \approx 0.5$-$0.7$; such low values are inconsistent with the time variable model which predicts $v_p/v_j = 1$, but would be produced by the K-H instabilities. On the other hand, Reipurth and Heathcote [5], using a slightly different determination of $\theta$ (=$30^o$), found $v_p/v_j \approx 1$, consistent with the time variable scenario. So, it seems that at this time the determination of the orientation angle of the outflow is too uncertain to be used as a way to decide for one or other model.

Using the same scenario of time-variability one can also explain the formation of the multiple bow shock structures observed at the heads of some YSO jets. To investigate this case, Gouveia Dal Pino and Benz [39] performed a simulation with long variablity period ($t>>\tau_{dy}$). The simulation produced a jet with a pair of bow shocks separated by a trail almost starved of gas extending for many jet radii in agreement with the observations. The leading bow shock or working surface developed at the head of the jet has a similar structure to that of steady state jets (see Fig. 7). The bow shock produced from the second ejection propagates downstream on the diffuse tail behind the leading working surface. It has a double shock structure similar to the one of the leading working surface and on average propagates faster. The postshock radiative material also cools and forms a cold shell whose density is much smaller than that of the shell of the leading working surface. This result is consistent with observations that suggest that the emission of the internal working surface is of lower intensity and excitation than that of the leading working surface.

All the proposed models above predict that the line emission in the knots is produced by shock-heated gas in the jet flow. Recently, however, a preliminary study by Bacciotti et al. [19], based on the diagnostics of low ionization state of the jet material (with an estimated ratio of the electron-to-hydrogen denstiy $x = n_e/n_H \approx 0.1$), suggests that the observed jet emission could be the product of soft compressions of the central portion of the flow, caused by "damped" K-H instabilities. In this picture, the emission would be connected to the heating produced not by shocks but instead by the mild compressions. Analytical and numerical calculations are still required to support or not this idea.

Finally, the possibility of the formation of the emission knots as condensations driven by thermal instabilities has been discussed by Gouveia Dal Pino & Opher [44] (see also e.g., [45-49] for a discussion on thermal instabilities in AGN jets and other astrophysical sources). However, while this mechanism may play a role in the formation of the filaments observed in AGN jets and lobes, it seems to be improbable in the case of the YSO jets as the thermal unstable perturbations in this case are overwhelmed by dynamical effects along the cooling beam, as indicated by numerical simulations [50].

## 5. MOMENTUM TRANSFER BETWEEN JET AND ENVIRONMENT

It is well known that many young stellar objects are associated not only to the highly collimated, fast YSO optical jets (100-500 km/s) investigated above, but also to less-collimated, slower molecular outflows (< 20 km/s) detected at radio wavelengths. The correlation between both kinds of outflows suggested since the beginning a "unified model" in which the optical jet entrains the molecular gas of the surrounding environment driving the molecular outflow [e.g., 51]. Recent determinations of the momentum carried by YSO jet have indicated that they carry, in fact, enough momentum to drive the associated molecular outflows [e.g., 19]. For example, taking a number density of ions and atoms $n \approx 1.1 \cdot 10^4$ cm$^{-3}$ in HH 34 jet and $n \approx 9 \cdot 10^3$ cm$^{-3}$ in HH 111 jet, Bacciotti et al. [19] estimated momentum rates $dP/dt = \pi R_j^2 \rho_j v_j^2 \geq 6.7 \cdot 10^{-5}$ M$_\odot$ yr$^{-1}$ km s$^{-1}$ for HH 34 jet, and $dP/dt \geq 1.3 \cdot 10^{-4}$ M$_\odot$ yr$^{-1}$ km s$^{-1}$ for HH 111 jet, which are larger than the estimated momentum rates for the associated molecular and neutral outflows: $\approx 5 \cdot 10^{-5}$ M$_\odot$ yr$^{-1}$ km s$^{-1}$ and $\approx 1.7 \cdot 10^{-5}$ M$_\odot$ yr$^{-1}$ km s$^{-1}$ for the HH 111 and 34 associated outflows, respectively. These determinations have motivated several authors to explore the mechanisms by which the YSO jets can transfer momentum to the molecular outflows [23, 39, 52-59].

Two basic processes of momentum transfer between a jet and a quiescent ambient medium can be distinguished: (i) the prompt entrainment, which is the transfer of momentum through the bow shock at the head of the jet; and (ii) the lateral (or turbulent) entrainment which refers to ambient gas that is entrained along the sides of the jet. This type of entrainment is the result of turbulent mixing through K-H instabilities at the interface between the jet and the ambient medium. A description of the dynamical properties of the lateral entrainment process in mixing layers can be found in [58-59] and the dynamics of the entrainment at the head of a jet has been investigated analytically in [54-55]. In the last case, the momentum transfer efficiency through a bow shock, $\varepsilon_{bs}$, which is given by the ratio between the rate of transfer of momentum to the ambient medium through the bow shock and the total input rate of momentum at the jet inlet, can be evaluated analytically and is given by [e.g., 23]:

$$\varepsilon_{bs} \approx 1/\alpha \, [ \, 1 + (\eta\alpha)]^{2} \qquad (7)$$

where $\eta$ and $\alpha$ have been defined in eq. (1).

Investigating both mechanisms through 3-D numerical simulations of steady-state jets moving in an ambient with Mach numbers in the range $3 < M_a = v_j/c_a < 110$ and density ratios between the jet and the ambient medium $0.3 < \eta < 100$, Chernin et al. [23] find that the lateral (turbulent) entrainment along the jet beam is important only for low Mach number ($M_a \leq 6$), low density ratio ($\eta \leq 3$) jets. Although this mechanism can be relevant for AGN jets which fit those values of $M_j$ and $\eta$, it is not the case for the typical YSO jets for which $10 < M_j < 40$ and $\eta \geq 1$. These results are illustrated in Fig. 8. The momentum transfer efficiency found in their numerical models is in agreement with eq. 7 for high Mach number jets indicating that in these cases the entrainment at the bow shock is dominant. So, YSO jets seem to transfer their momentum predominantly at the bow shock. These results rule out models which propose that molecular outflows are produced through turbulent mixing [52, 53, 59].

Chernin et al. [23] also find that the extremely high velocity (EHV) CO features observed at the molecular outflows [e.g., 60] can be formed in the swept-up post-shock gas behind the bow shock. These results are confirmed by the more recent numerical calculations of Raga et al. [58] who computed intensity maps of the molecular emission produced at the head of a jet traveling into a molecular environment and find that a trail of molecular material is left behind the passage of the bow shock.

The effect on the environment of the passage of successive bow shocks (or internal working surfaces) of a time variable jet (like those discussed in Section 4) has also been investigated. Raga and Cabrit [55] proposed that the molecular emission could be identified with the swept-up environmental gas that fills the cavity formed behind the internal working surfaces of a time-dependent ejected jet. The simulations of intermittent radiative cooling jets with small ($\tau \approx \tau_{dy} = R_j/v_j$) variability period discussed in section 4 (see Fig. 6) show no evidence that the ambient material pushed aside by the internal bow shocks have had time to fill the cavities behind them [39]. The swept-up ambient material cools at the edge of the cavities. But, the jet with long variability period ($\tau > \tau_{dy}$) depicted in Fig. 7, clearly shows that the ambient gas swept up by the leading working surface eventually refills the cavity formed behind it [39], in agreement with Raga and Cabrit's model.

Also, Raga et al. [56] explored a model in which the molecular outflows correspond to the turbulent envelopes of mixed jet and ambient material developed around time-dependent jets. According to the observations, the molecular gas can typically extend over transverse sizes which are about 20 times wider than the jet diameter. The results for the evolved system of Fig. 7 indicate that the envelope does not exceed ~6 jet diameters. However, this "narrowing" tendency of the distribuition of the surrounding gas in the simulation can be a consequence of the assumed periodic boundaries on the transverse peripheries of the ambient computational domain [39].

## 6. ENVIRONMENTAL EFFECTS

Jets emerging from YSOs and AGNs propagate into complex ambients [61-62]. YSO jets, in particular, are immersed in molecular clouds which are generally non-homogeneous and formed of many smaller, dense clouds. Basically all the models discussed so far have considered the propagation of the jets in initially homogeneous ambients. We now make a brief description of recent numerical results obtained for cooling steady-state jets propagating into more realistic environments [63-64]. In these models it is assumed an initial isothermal ambient medium (with a temperature $T_a = 10^4$ K) with a power-law density (and pressure) distribution, $n_a(x) \propto x^\beta$, where $n_a$ is the ambient number density, x the longitudinal jet axis, and $\beta = \pm 5/3$ for positive and negative density gradients, respectively. Such profiles are consistent with the observed density distribution of the clouds which involve protostars [65]. A negative density gradient atmosphere may represent, for example, the ambient that a jet finds when it emerges from the parent source and propagates through the cloud that involves the system. A positive density gradient atmosphere can be found when a traveling jet strikes an external cloud. The density of the cloud increases as the jet submerges into it.

As an example, Fig. 9 depicts a jet propagating into an ambient medium of increasing density (pressure). In these cases, the jets have their cocoon compressed and pushed backwards by the ambient ram pressure and the beam is highly collimated. The compressing medium induces the development of Kelvin-Helmholtz instabilities which cause the beam focusing, twisting and flapping and the formation of internal knots close to the head. The bow shock-like structure that appears at the head of jets propagating into homogeneous ambients (see Fig. 4), is substituted by a thin nose cone in the jet of Fig. 9. This morphology remarkably resembles some observed YSOs jets such as HH 30 [66] which has a wiggling nose cone structure with no bow-shock like feature and a chain of emission knots far from the source very similar to that of Fig. 9 (other examples include HH83 and HH 110). Biro et al. [67] have performed numerical calculations of a jet ejected from a precessing source into a homogeneous ambient which also produced a wiggling beam.. In their case, however,

**Fig. 8** The trajectories of the particles of the ambient medium. The top figure shows the case of a jet with η = 1and $M_a$ =1.1 and the bottom figure the case of a jet with η=10 and $M_a$ =10.   The turbulent eddies in Fig. 8b are a signature of the turbulent entrainment. In Fig. 8a, the˝straight trajectories of the ambient particles indicate that the prompt momentum transfer through the bow shock is the dominant process in this case (from Chernin et al. 1994 [23]).

**Fig. 9** Central density contours of a jet propagating into an ambient medium with positive density gradient. The initial conditions are: $\eta(x_o)=1$, $n_a(x_o)=200$ cm$^{-3}$, $R_j=2\times10^{15}$ cm, $v_j=398$ km/s, $M_a=24$, $q_{bs}\approx8.1$, $q_{js}\approx0.3$, and $R_j/c_a=38.2$ yrs. The coordinates x and z are in units of $R_j$. The contour lines are separated by a factor of 1.3 and the density scales from $\approx 0$ up to $800/n_a$. (from Gouveia Dal Pino, Birkinshaw, and Benz 1994 [63]).

**Fig. 10** Central density contours and velocity distribution of a jet propagating into an ambient medium with positive density gradient. The initial conditions are the same of Fig. 9 (from Gouveia Dal Pino, Birkinshaw, and Benz 1994 [63]).

the mechanism producing such an effect is the "garden-hose" instability".

An example of a jet propagating in an ambient medium with negative density gradient is shown in Fig. 10. Negative density ambients produce broad and relaxed jets, with no knots and only faint radiative shocks at the head with poor emission. Therefore, they may be important in the invisible portions of the observed outflows.

Compared to a cooling jet, an adiabatic jet propagating into similar stratified ambients is slightly less affected by the stratification effects due to its higher pressure cocoon which better preserves the structure of the beam. This result may be particularly applicable to extragalactic jets.

## 7. THE ORIGIN OF THE JETS

Although considerable progress has been made toward understanding the jet structure and propagation, no consensus has been reached concerning the basic mechanism for its origin. The measured velocities and mass discharges ($dM/dt \approx 10^{-8}$-$10^{-7}$ $M_\odot$ yr$^{-1}$) in the YSO jets rule out thermal or radiation pressure as the driving mechanisms [e.g., 68]. Alternative possibilities include: (i) magnetic-pressure-driven outflows; or (ii) magnetocentrifugally-driven outflows (with magnetic tapping of the rotational energy of the stellar object or of its circumstellar disk, or of both). There now exists strong evidence for the presence of nearly Keplerian accreting disks around YSOs and observational support of their correlation with the bipolar outflows [e.g., 69]. Also, polarization measurements indicate that the magnetic field of the molecular cloud cores are preferentially aligned with the symmetry axes of the embedded circumstellar disks and the associated bipolar flows [70].

In the first class of models mentioned above [71-75] the outflow material is accelerated from the surface of the circumstellar disk by magnetic pressure gradients. In these models the magnetic field at the top of the disk has a strong azimuthal component and the acceleration is due to its uncoiling like a spring above the surface. In particular, Pringle [72] proposes that the strong azimuthal field arises in the boundary layer between the accretion disk and the central star. In this boundary layer, shearing takes place and the azimuthal component of the field is amplified until it can diffuse out of the boundary layer region. Regions of strong field are expected then to decouple from the gas through the Parker instability, so that the Alfvén speed rapidly becomes larger than the escape velocity and an outflow emerges along the symmetry axis. However, these models are, in general, unsteady. In Pringle's scenario, for example, the large axial (z) gradients of the azimuthal field ($B_\phi$) that are postulated to exist at the base of the outflow can be possibly established only by transient events.

Among the magnetocentrifugal models we may distinguish those in which the outflow arises from the stellar surface (pure stellar outflows), those in which it originates on the accretion disk (pure disk outflows), and those in which the outflow arises from an interacting star-disk system. These models seem to be more attractive than the magnetic pressure models discussed above because they provide a very simple and direct method for transferring angular momentum excess out of the system, thus, slowing down the protostar as required by the observations (see below), without having to invoke poorly undesrtood viscosity processes [e.g., 1].

Magnetocentrifugal models of pure disk outflows have been worked out in some detail [70, 76-79], Wardle and Königl [70] construct explicit solutions of the vertical disk structure. They assume a disk threaded by a large scale, open magnetic field. Most of the material lies in a sub-Keplerian, quasi-hydrostatic layer close to the midplane (z=0, expressed in cylindrical coordinates). The field is coupled to the weakly ionized disk material by ion-neutral collisions that cause the inward slip of the neutral gas relative to the ions and field at a velocity $v_d$ (ambipolar diffusion). The ions lag the neutrals because of the frictional drag. The back reaction of the ions on the neutrals causes neutral gas to loose angular momentum to the field. This in turn enables the neutral to drift toward the center and exert a radial drag of the field lines. The drag must be balanced by magnetic tension, so the field lines bend outward from the rotation axis. At higher latitudes, the field dominates the energy density. The azimuthal velocity of the field increases with the height and eventually the field lines overtake the neutral fluid and start to transfer angular momentum back to matter and to accelerate it centrifugally into a collimated wind. While attractive, the pure-disk wind model applied to YSOs has some limitations [80]: (i) the large radial drift needed in the midplane (enforced by rotation at sub-Keplerian speeds) at radius $r \approx 10^{15}$ cm (where observed disks are believed to end) leads to accretion timescales $r/v_d$ about three orders of magnitude shorter than the values, $10^6$ - $10^7$ yr, inferred for YSO disks [81]; (ii) ohmic dissipation and ambipolar diffusion will possibly rid the disk interior of magnetic fields with the intensities required in the disk-outflow models for all radius $r \approx (0.1 - 20) \, 10^{13}$ cm. At small radius, heating by stellar photons and viscous accretion provide a good degree of ionization. Thus, effective mechanical coupling between the gas and the magnetic fields will possibly exist only in the inner regions of the disk.

In the context of magnetocentrifugal pure stellar outflow models, Hartmann and MacGregor [82] proposed that the material would stream off a contracting rotating protostar along radial magnetic field lines rooted in the star. Most of the mass loss was supposed to occur in the equatorial plane, but diverted towards the poles by magnetic pressure gradient in the azimuthal field. Shu et al. [83], modified this model to incorporate a more realistic scenario including an accretion disk. In this model the neutral molecular outflows (see Section 5) represent gas transferred from the accretion disk onto the magnetic field lines of the YSO and launched out from two narrow equatorial bands that straddle the disk, while the ionized jet originates as a normal stellar wind near the poles of the protostar. A potential difficulty with this model is that it requires the star to be rotating at high angular velocities close to breakup, which appears to be inconsistent with observations of optically revealed young stars which usually rotate at speeds an order of magnitude below breakup. In more recent work Shu et al. [80, 84-85] proposed a generalization of this previous model in order to surpass its difficulty. They assumed a viscous and imperfectly conducting disk (with larger magnetic diffusivity $\eta_D$) accreting steadily (with smaller accretion rate $dM_D/dt$) onto a young star with a strong magnetic field. For an aligned stellar magnetosphere, shielding currents in the surface layers of the disk will prevent stellar field lines from penetrating the disk everywhere except for a range of radii around $r = R_X$, where the Keplerian angular speed of rotation $\Omega_X$ equals the angular speed of the star $\Omega_S$. For the low disk accretion rate and high magnetic fields associated with typical young stars, $R_X$ exceeds the radius of the star $R_S$ by a factor of a few, and the inner disk region is effectively truncated at a radius $R_t < R_X$. Between $R_t$ and $R_X$, the closed field lines bow sufficiently inward and the accreting gas attaches to the field and is funneled dynamically down the effective potential (gravitational plus centrifugal) onto the star. The associated magnetic torques to this accreting gas may transfer angular momentum mostly to the disk. Thus, the star can spin slowly as long as $R_X$ remains $> R_S$. For $r > R_X$ field lines threading the disk bow outward making the gas off the midplane rotate at super-Keplerian velocities. This combination drives a magnetocentrifugal wind with a mass-loss rate $dM_W/dt$ equal to a definite fraction f of the disk accretion rate $dM_D/dt$. For high disk accretion rates, $R_X$ is forced down to the stellar surface, the star is spun to breakup, and the outflow is generated in an identical manner to that proposed in the previous model of Shu et al. [83]. A schematic diagram of this model is presented in Fig. 11. A quantitative description of this model is presented in [84-85]. They find that stellar magnetic fields of $\approx$ few kG can drive outflows with mass-loss rates of $10^{-6}$ $M_\odot$ yr$^{-1}$ from rapidly accreting YSOs (rotating near breakup) and $10^{-8}$ $M_\odot$ yr$^{-1}$ from slowly accreting (slowly rotating) young stars. The mechanism described can accelerate outflows from these systems to velocities $\approx$ few 100 km yr$^{-1}$ within few stellar radii $\leq 10$ $R_S$.

Observational tests of the models above are very difficulty because the central sources and the associated accretion disks (which lie in a region of $\approx 10^{-3}$ pc) are not well resolved by the instruments. So, at this stage we cannot rule out the magnetic pressure models, although the magnetocentrifugal mechanisms (and, in particular, the unified star-disk model) appear to offer a more natural scenario for explaining the formation of collimated outflows. It is also possible that all these models play a role at different stages of the star-disk evolution.

More generally, these mechanisms are believed to be universal in the production of bipolar outflows from a wide range of astrophysical objetcs [e.g., 70, 75, 80]. In the case of the AGN jets, also there is now evidence for the presence of dusty molecular disks near the central nucleus [86] which gives support to disk-driven outflow models or the combined source-disk-driven outflow models.

Since the outflows are in principle capable of transporting the excess of angular momentum of the accreting matter as well as most of the gravitational energy that is liberated, one may attribute the ubiquity of bipolar outflows to the fact that the winds are a necessary ingredient in the accretion process in that they carry away the angular momentum that needs to be removed in order for the accretion to proceed [78].

## 8. SUMMARY

Supersonic collimated outflows are a common phenomenum in the universe and span a large range of intensities, from the powerful examples observed to emerge from the nuclei of active galaxies (or AGNs) to the jets associated to low-mass Young Stellar Objects (YSOs) within our own Galaxy. Thanks to their proximity, YSO jets constitutes the best laboratories for cosmic jet investigation. This paper tried to review the main properties of these objects and the proposed models for their origin and structure. Possible applications of the theories to the AGN jets were also briefly discussed.

**Fig. 11** Schematic diagram of the principal components of a magnetocentrifugally driven outflow from a disk-star system. Dotted curves correspond to the loci where the gas makes a transition from subsonic (or sub-Alfvénic) to supersonic (or super-Alfvénic) flow. Matter interior to the radius $R_x$ diffuses onto field lines that bow inward and is funneled onto the star. Matter at $r > R_x$ diffuses onto field lines that bow outward and launch a magnetocentrifugally driven wind. These lines (which are opened by the ram pressure of the outflowing gas) are fixed in place at the equatorial plane by a combination of the inward press of the accretion flow and the outward centrifugal press of the field lines still closed on the star and embedded in matter forced to rotate at super-Keplerian speeds. Between the t-region (where angular momentum flows into disk from the funnel flow) and the x-region (where angular momentum flows out of the disk into the wind) lies a dead zone where the matter diffuses quasi-statically across field lines. It is in pressure equilibrium with the funnel and wind flows (from Shu et al. [80]).

Simple one-dimensional analytical approach makes possible to model the basic features of high Mach number, heavy cooling jets like the typical YSO jets (Sections 3 and 4). A supersonic jet propagating into a stationary ambient gas will develop a shock pattern at its head or working surface. The impacted ambient material is accelerated by a forward bow shock and the jet˝material is decelerated at a shock wave as it reaches the head, i.e., the jet shock. Numerical hydrodynamical simulations of jets propagating in initially homogeneous ambient medium confirm the analytical predictions and reveal the development of a dense shell at the head of the jet which is formed by the cooling of the shock-heated material at the working surface. The shell eventually becomes dynamically and thermally unstable and fragments into pieces which resemble the clumpy structure of observed HH objects at the heads of YSO jets˝(Section 3).

The chain of aligned knots observed in typical YSO jets is probably mainly formed by time-variability in the flow ejection velocity (Section 4). Supersonic velocity variations (with periods close to the transverse dynamical timescale $\tau_{dy} = R_j/v_j$) quickly evolve to form a chain of regularly spaced radiative shocks along the jet with large proper motions and low intensity spectra in agreement with the observed properties of the knots. The knots widen and fade as they propagate downstream and eventually disappear. So, this mechanism favors the formation of knots closer to the driving source in agreement with typical observations. Longer variability periods ($t \gg \tau_{dy}$) can also explain the multiple bow shocks observed in some jets. From these results we can make the conjecture that, possibly, the superluminal knots which are observed to emerge form some AGN jets in the parsec (pc) scales [e.g., 11-12] could be also the product of time-variability in the ejection mechanism from the corresponding central sources.

Shocks driven by Kelvin-Helmholtz (K-H) instability are viable candidates for knot production in jets for which the adiabatic approach is good (AGN jets). For radiative cooling jets (YSO jets), the cooling reduces the pressure in the cocoon that surrounds the beam. As a result, the cocoon has less pressure to drive K-H instabilities and thus, reflect internal shocks in the beam. In the numerical simulations of jets propagating continually into homogeneous ambient medium, only very heavy cooling jets ($\eta \approx 10$) shows the formation of internal shocks. However, an efficient formation of a "wiggling" chain of knots, primarily close to the jet head, by K-H instabilities is found in simulations of jets propagating into environments of increasing density (and pressure) (Section 6). In this case the cocoon is compressed and pushed backwards by the ambient ram pressure and the beam is highly collimated. The compressing medium then induces the development of the K-H instabilities. This structure remarkably resembles some observed YSO jets (e.g., HH 30 jet) which do show a wiggling structure with knot formation closer to the jet head.

Ambient medium with negative density gradient, on the other hand, produces a broad and relaxed jet with no internal knots. The shocks at the head are faint and poor in emission. Thus, this scenario may apply to the invisible portions of some observed outflows (Section 6).

YSO jets appear to carry enough momentum to drive the observed less collimated slower molecular outflows that are also associated to YSOs (Section 5). Numerical simulations indicate that for YSO jets the transfer of momentum to the molecular ambient medium occurs predominantly through the bow shocks and not by turbulent mixing through the lateral discontinuity layer between jet and ambient medium. (This last mechanism is more relevant in low Mach numbers ($M_j \leq 3$), light ($\eta \leq 3$) jets and thus, more appropriate in AGN jets.) The molecular emission could then be possibly identified with the swept-up environmental gas by the passage of the internal working surfaces (or bow shocks) of a time-dependent ejected jet.

Theoretical models for the origin of the jets involve wind acceleration from the central YSOs or from their accreting circumstellar disks (Section 7). Basically, two viable mechanisms for driving the outflows from the sources have been investigated in detail: (i) outflow acceleration by magnetic pressure gradients; and (ii) magnetocentrifugal acceleration. Among the magnetocentrifugal models we may distinguish those in which the outflow arises from a magnetized rapid rotating protostar, or from a rotating disk with substantial poloidal magnetic fields of open geometry, or a combination of a strongly magnetized protostar with an adjoining Keplerian disk. The magnetocentrifugal models seem to be more attractive than the magnetic pressure models as they provide a very simple and direct method for transferring angular momentum excess out of the system, thus, slowing down the protostar as required, without having to invoke poorly undesrtood viscosity processes. Observational tests to check the viability of these models, however, are very difficulty because the central sources and the associated accretion disks are not observationally resolved. Also, it is not impossible, that each of these models may play a role at different stages of the star-disk evolution. These mechanisms are believed to be universal in the production of bipolar outflows from a wide range of objetcs, ranging from the YSOs to the AGNs.

While of fundamental importance in the production and initial collimation of the jets (Section 7), magnetic fields have been neglected in most of the analytical and numerical modeling of the structure of the YSO jets since the observational estimates of their intensity suggest that magnetic fields are not dynamically important along the flow. However, they can become relevant once they are amplified by compression behind the shocks. Future work should then incorporate magnetic fields in the investigation of the YSO jet structure as it has already been done for AGN jets [e.g., 14].

## REFERENCES

1. Padman, R., Lasenby, A. N., and Green, D. A 1991, in: Beams and Jets in Astrophysics, P. A. Hughes (ed.), Cambridge Univ. Press, p. 484.